\documentclass[aps,prl,twocolumn,preprintnumbers,amsmath,amssymb,superscriptaddress]{revtex4}

\usepackage{graphicx}
\usepackage{pifont} 
\usepackage{wasysym} 

\begin{document}

\title{Coherent Josephson qubit suitable for scalable quantum integrated circuits}

\author{R. Barends}
\thanks{These authors contributed equally to this work.}
\author{J. Kelly}
\thanks{These authors contributed equally to this work.}
\author{A. Megrant}
\author{D. Sank}
\author{E. Jeffrey}
\author{Y. Chen}
\author{Y. Yin}
\altaffiliation[Present address: ]{Department of Physics, Zhejiang
University, Hangzhou 310027, China}
\author{B. Chiaro}
\author{J. Mutus}
\author{C. Neill}
\author{P. O'Malley}
\author{P. Roushan}
\author{J. Wenner}
\author{T. C. White}
\author{A. N. Cleland}
\author{John M. Martinis}
\affiliation{Department of Physics, University of California, Santa
Barbara, CA 93106, USA}

\date{\today}

\begin{abstract}
We demonstrate a planar, tunable superconducting qubit with energy
relaxation times up to 44~$\mu$s. This is achieved by using a
geometry designed to both minimize radiative loss and reduce coupling
to materials-related defects. At these levels of coherence, we find a
fine structure in the qubit energy lifetime as a function of
frequency, indicating the presence of a sparse population of
incoherent, weakly coupled two-level defects. This is supported by a
model analysis as well as experimental variations in the geometry.
Our `Xmon' qubit combines facile fabrication, straightforward
connectivity, fast control, and long coherence, opening a viable
route to constructing a chip-based quantum computer.
\end{abstract}

\maketitle

One of the outstanding challenges in building a quantum computer is
to balance coherence, connectivity and control in the qubits.
Superconductivity provides an appealing platform because it allows
for scalability: the conduction electrons condense into a macroscopic
quantum state, and large quantum integrated circuits can be made with
many elements having individual control lines. However, quantum
coherence in superconducting circuits has proven to be very delicate,
as it is easily disturbed by material defects, electron system
excitations, and radiative coupling to external wiring
\cite{martinis2005,grabovskij,catelani,wenner2013,riste,shaw,lafarge1993}.
To minimize these and other effects, many groups have recently begun
embedding qubits in three-dimensional superconducting cavities. These
3D qubits show high coherence, with energy relaxation times in 3D
transmon qubits between 30 and 140~$\mu$s
\cite{paik2011,rigetti2012}.

Here, we demonstrate a new design for a fully planar superconducting
qubit, based on the planar transmon \cite{koch,houck2007}, with
energy coherence times in excess of 40~$\mu$s. Our approach balances
coherence, connectivity, as well as fast control. We design the qubit
with high-quality coplanar waveguide capacitors, motivated by the
recent advances with superconducting resonators, yielding a modular
design with straightforward connectivity. The qubits are
frequency-tunable, which allows the implementation of fast two-qubit
gates: a controlled-Z gate \cite{strauch,dicarlo2009,mariantoni} can
then be implemented with high fidelity in 25~ns \cite{ghosh}. With
the coherence time exceeding single and two-qubit gates times by
three orders of magnitude, we believe that our device provides a key
ingredient for implementing a surface code quantum computer
\cite{fowler}.

We also find that the small remnant decoherence in these qubits comes
from a sparse bath of weakly coupled, incoherent defects. These
defects are clearly visible in the measured quantum time-resolved
spectroscopy, and give rise to frequency-dependent variations in the
energy relaxation rate. These results may also explain the variations
observed in lifetimes of 3D transmon qubits. We elucidate this
physics and improve the coherence by varying the capacitor geometry.

\begin{figure}[b]
    \centering
    \includegraphics[width=0.48\textwidth]{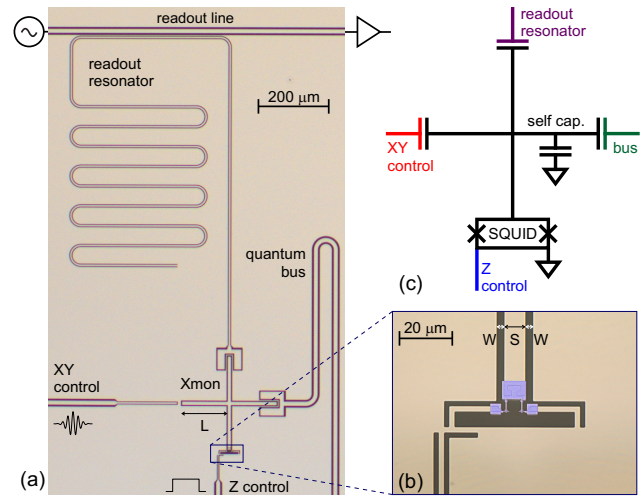}
    \caption{(Color online) (a) Optical micrograph of the planar `Xmon' qubit, formed by the Al superconducting film (light) and the exposed sapphire substrate (dark).
    The qubit is capacitively coupled to a quarter wave readout resonator (top), a quantum bus resonator (right),
    a XY control line (left), and inductively coupled to a Z control line (bottom). The Xmon arm length is $L$.
    (b) The inset shows the shadow evaporated Al junction layer in false color (blue). The junction size is $0.30\times0.20~\mu$m$^2$.
    The capacitor central linewidth is $S$, the gap width is $W$.
    (c) The electrical circuit of the qubit.}
    \label{fig:xmon}
\end{figure}

Our device is shown in Fig.~\ref{fig:xmon}a, formed by patterning the
Al metal (light), exposing the sapphire substrate (dark). The qubit
is the cross-shaped device. Its four arms connect to separate
elements, each having a different function: a coplanar waveguide
resonator for readout on the top, a quantum bus resonator on the
right to mediate coupling to other qubits, XY control on the left to
excite the qubit state, and Z control on the bottom to tune the qubit
frequency. The cross is the qubit capacitor, which connects at the
bottom to the tunable Josephson junction, formed by the rectangular
ring-shaped superconducting quantum interference device (SQUID), see
Fig.~\ref{fig:xmon}b. The rectangular ring is intersected by two
identically sized Al tunnel junctions (blue in Fig.~\ref{fig:xmon}b).
The electrical circuit is equivalent to that of a grounded transmon
\cite{koch}, with the capacitor in parallel with the tunable junction
(Fig.~\ref{fig:xmon}c). In a clear departure from the traditional
floating transmon with an interdigitated capacitor (IDC)
\cite{houck2007}, we chose to form the qubit capacitor by
intersecting two coplanar waveguide lines.

In prior work, we showed that highly coherent coplanar waveguide
resonators can be fabricated, having quality factors of about
$1.5\times10^6$ at the single photon occupation level. These
resonators were made from molecular beam epitaxy (MBE) Al on
oxygen-cleaned sapphire \cite{megrant2012}. This shows that a
straightforward path to high coherence comes from a combination of I)
MBE Al as high quality material, II) coplanar waveguides having low
radiative loss, and III) embedding in a groundplane. We therefore
embed the qubit in an uninterrupted groundplane, with thin Al lines
at the capacitor ends tying the groundplanes together; this
suppresses parasitic slotline modes in the control lines and
resonators as well.

Connectivity is accomplished by coupling each of the qubit's arms to
a distinct element with specific functionality. Three of the
connections are easily made with a coupling capacitor, as the qubit
is connected to ground. An advantage of this approach is that each
coupling can be individually tuned and optimized. To this end, we
have also separated out qubit control. The XY control drive line is
connected with a coupling $C_c = 60~$aF, which allows us to excite
the qubit state in 10~ns but hardly affects coherence, with an
estimated $T_1$ of 0.3~ms. The Z control also combines speed and
coherence. The drive line is galvanically connected to the SQUID to
allow for a large inductive coupling with a mutual inductance of
$M=2.2~$pH. We are able to rapidly detune the qubit on the timescale
of a nanosecond \cite{supmat}. The measured parasitic coupling
between the Z line and the qubit gives an estimated $T_1$ of
$\sim$30~ms \cite{macc}.

The qubit design outlined above balances coherence, control and
connectivity in the qubit, using a robust design for the capacitor
and a modular separation of functionality. Our approach differs on
key points from previous implementations. With this experimental
nature in mind, we name our qubit the `Xmon'. While the cross-shaped
qubit capacitor may emphasize this name, more arms can be added to
allow for more connectivity.

\begin{figure}[t]
    \centering
    \includegraphics[width=0.48\textwidth]{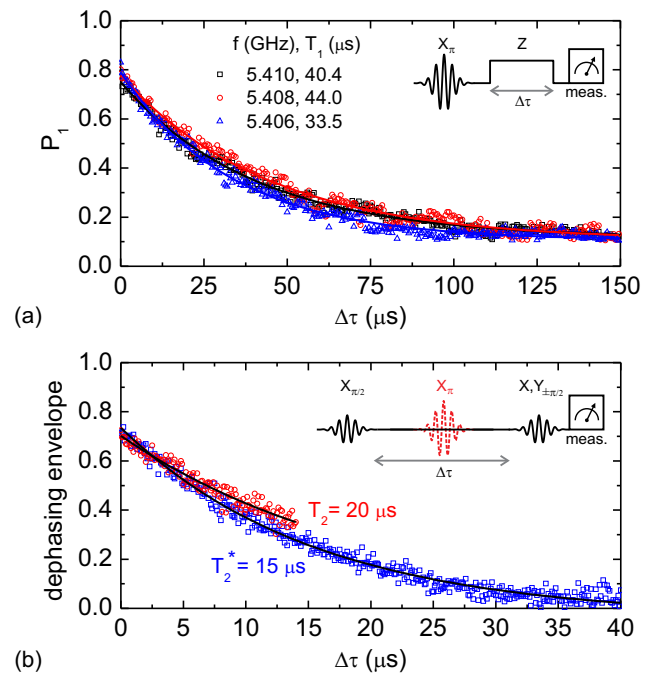}
    \caption{(Color online) (a) Qubit energy decay at three nearby frequencies.
    The qubit frequency is adjusted by applying a rectangular pulse with length $\Delta \tau$ on the Z line.
    The pulse sequence is shown in the inset.
    (b) Ramsey $T_2^*$ and spin echo $T_2$ dephasing envelopes at the flux insensitive point, measured by phase tomography.
    The inset shows the pulse sequence; for the spin echo we apply a refocussing pulse (dashed).
    We apply four phases to the last pulse for phase tomography, to measure the decay envelope.
    Spin echo measurements are limited by electronics to 14~$\mu$s.
    The energy coherence for the dephasing measurement was $T_1=18~\mu$s.}
    \label{fig:xmoncoherence}
\end{figure}

We find a dramatic increase in Xmon energy coherence compared to the
traditional planar transmon, measuring decay times up to
$T_1=44~\mu$s, see Fig.~\ref{fig:xmoncoherence}a. The qubit $T_1$ is
measured by exciting it with a $\pi$-pulse, and measuring its state
after a variable time $\Delta \tau$. We find that the excited state
probability decays exponentially. We find Ramsey and spin echo phase
coherence times up to $T_2^*=15$~$\mu$s and $T_2=20$~$\mu$s at the
flux insensitive point, respectively, see Fig.
\ref{fig:xmoncoherence}b. The energy decay at this point is measured
to be $T_1=18~\mu$s. The dephasing envelopes are measured using
tomography. The first pulse is $X_{\pi/2}$, followed by $X_{\pi/2}$,
$X_{-\pi/2}$, $Y_{\pi/2}$, or $Y_{-\pi/2}$ (see inset for the pulse
sequence), producing fringes with different phases. The dephasing
envelopes follow an exponential decay. As the limit of $T_2 = 2 T_1$
has not been reached \cite{koch}, this indicates the presence of an
additional dephasing channel. This channel, as well as dephasing away
from the flux insensitive point, is presently under investigation.

The qubits used here had ground to excited state transition
frequencies around 6~GHz when unbiased, nonlinearities around
230~MHz, and a ratio of Josephson to charging energy $E_J/E_C\sim95$.
We employ a dispersive, high-power single-shot readout scheme with a
70-85\% fidelity \cite{reed2010}. The readout resonator frequencies
used are 6.4-6.7~GHz, the loaded quality factor is $Q_l=10^4$, and
the resonator-qubit coupling strength is approximately 40~MHz.
Measurements were done in a dilution refrigerator with a base
temperature of 30~mK, with multistage infrared shielding
\cite{barends2011}. Magnetic fields were reduced by room temperature
and cryogenic magnetic shields, with nonmagnetic microwave connectors
\cite{ezform}.

The results in Fig. \ref{fig:xmoncoherence} show that tunable
superconducting qubits with a planar geometry can have $T_1$ values
in excess of 40$~\mu$s. In fact, this $T_1$ corresponds to the MBE Al
resonator quality factors \cite{megrant2012}, for which
$T_1=Q/\omega$ is also about 40$~\mu$s. The combination of long
energy and phase coherence times compare well with previously
reported values for planar superconducting Al qubits: for transmons
$T_1=9.7~\mu$s and $T_2^*=10~\mu$s \cite{chow}, for charge qubits
$T_1=200~\mu$s and $T_2^*=0.07~\mu$s \cite{kim2011}, for flux qubits
$T_1=12~\mu$s and $T_2^*=2.5~\mu$s \cite{bylander2011}, and for the
fluxonium $T_1=10~\mu$s and $T_2^*=2~\mu$s \cite{manucharyan2012}. In
fact, the Xmon approaches the long coherence found in 3D transmons
\cite{paik2011,rigetti2012}. Very recently, TiN planar devices have
shown long coherence \cite{chang,sage}, encouraging using Xmon
geometries with this material.

\begin{figure}[t]
    \centering
    \includegraphics[width=0.48\textwidth]{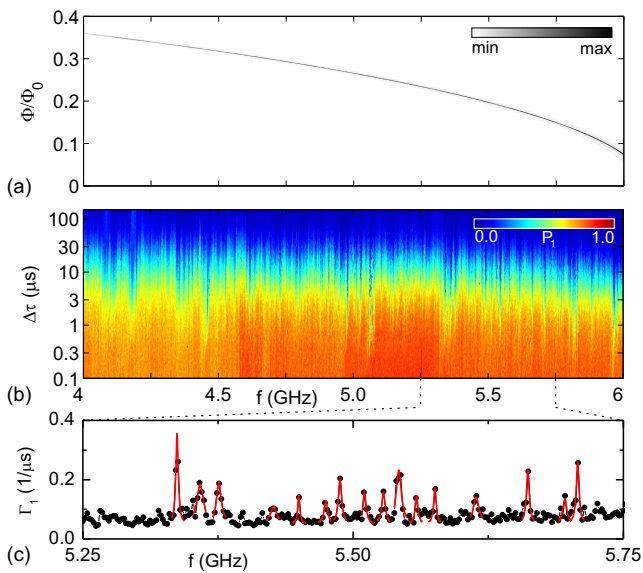}
    \caption{(Color online) (a) Qubit spectroscopy for device with $S$,$W$=8,8~$\mu$m.
    A smooth curve is formed by the high transmission (grey), measured on resonance with the readout resonator,
    which indicates when the qubit is excited.
    (b) Swap spectroscopy of the same qubit. The qubit is detuned from 4 to 6~GHz (stepsize 2~MHz),
    and the delay time is varied from 100~ns to 150~$\mu$s. See the inset of Fig. \ref{fig:xmoncoherence}a for the pulse sequence.
    (c) Qubit relaxation rate at qubit frequencies from 5.25 to 5.75~GHz, extracted from the data in (b). Sharp peaks above a background are observed.
    The peaks are fitted to Eq.~\ref{eq:incoherentdecay} (solid lines).}
    \label{fig:swapspec}
\end{figure}

\begin{figure}[t]
    \centering
    \includegraphics[width=0.48\textwidth]{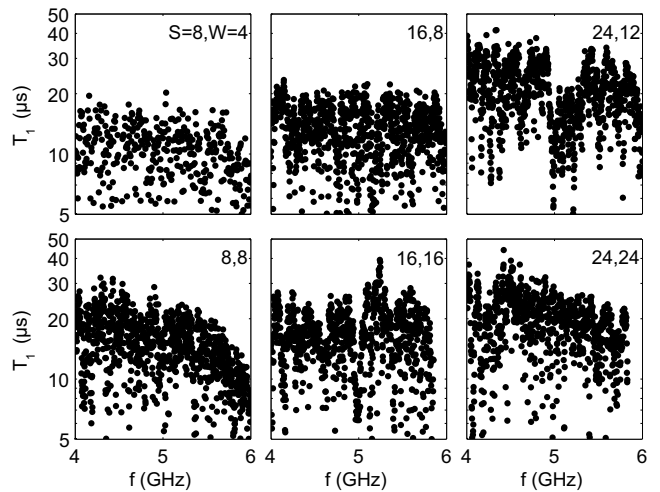}
    \caption{Frequency dependence of $T_1$ for six qubits with different $S$ and $W$ (see Table \ref{table:parameters}).
    The frequency stepsize is 5~MHz for $S$,$W$ = 8,4~$\mu$m and 2~MHz otherwise.
    See Supplementary Material \cite{supmat} for the corresponding decay rates.}
    \label{fig:t1}
\end{figure}

We find that the energy relaxation depends on qubit frequency, as we
observe different exponential decay rates for small changes in
frequency. As shown in Fig. \ref{fig:xmoncoherence}a, we find $T_1$
values from 34 to 44$~\mu$s in a 4~MHz band near 5.4~GHz. In order to
elucidate this further, we performed a spectroscopic scan on the
qubit, shown in Fig. \ref{fig:swapspec}a. The qubit frequency
displays the expected dependence on applied flux $\Phi$ \cite{koch},
varying smoothly without visible splittings, indicating that strongly
coupled defects, which manifest as avoided level crossings
\cite{martinis2005}, are virtually absent. We then performed a
quantum analogue of time-resolved spectroscopy (swap spectroscopy
\cite{cooper2004}), shown in Fig.~\ref{fig:swapspec}b. The
probability of the excited state (color) is plotted for $\Delta \tau$
from 100~ns to 150~$\mu$s (logarithmic vertical scale) and qubit
frequencies from 4 to 6 GHz. We find that the probability decays
exponentially, but with a fine structure of variable energy
relaxation, and distinct peaks in the energy decay rate
(Fig.~\ref{fig:swapspec}c). We do not observe any chevron
interference patterns \cite{cooper2004}, implying no defects interact
coherently with the qubit state. After cycling the temperature to
4.2~K the fine structure is altered, but the overall image remains
unchanged. We count approximately 30 regions with reduced coherence
($T_1<8$~$\mu$s) per GHz in Fig.~\ref{fig:swapspec}b.

\begin{table}[b!]
\centering \caption{Geometric parameters for the Xmon qubit
capacitors as defined in Fig.~\ref{fig:xmon} along with their
frequencies. Groups of three qubits indicate that the devices are on
the same chips.}
\begin{tabular}{|c|ccc|ccc|}
  \hline
  $S$ ($\mu$m) & 8 & 16 & 24 & 8 & 16 & 24 \\
  $W$ ($\mu$m) & 4 & 8 & 12 & 8 & 16 & 24 \\
  $L$ ($\mu$m) & 130 & 130 & 130 & 165 & 165 & 165 \\
  \hline
  $f_{10,\mathrm{max}}$ (GHz) & 6.094 & 6.158 & 6.071 & 6.080 & 5.883 & 5.846 \\
  nonlinearity (MHz) & 224 & 228 & 222 & 220 & 225 & 223 \\
  \hline
\end{tabular}
\label{table:parameters}
\end{table}

We explored the dependence of the qubit coherence time on capacitor
geometry, using six different qubit capacitor designs; the central
line width $S$, gap width $W$ and arm length $L$ were varied, while
the capacitance value \cite{ease} and junction parameters are kept
the same. The parameters are listed in Table \ref{table:parameters};
see Supplementary Material \cite{supmat} for a micrograph. We find
that the swap spectroscopy measurements of the different designs
share the same characteristics as shown in Fig.~\ref{fig:swapspec}b:
a fine structure with varying exponential decay. The energy
relaxation times extracted from the measurements are shown in
Fig.~\ref{fig:t1}. With larger capacitor size the overall energy
relaxation time increases: when changing $S$,$W$ from 8,4~$\mu$m to
16,8~$\mu$m and 24,12~$\mu$m, the $T_1$ improves from a band of
values between 8-15 to 10-20, and 20-40~$\mu$s, respectively.
Importantly, both the upper as well the lower bounds on $T_1$
increase with capacitor width. This is repeated in the qubits with
$S$,$W$ ranging from 8,8~$\mu$m to 16,16~$\mu$m and 24,24~$\mu$m. The
reduction of $T_1$ at frequencies approaching 6~GHz is due to Purcell
decay into the readout resonator \cite{houck2008}. We emphasize that
these $T_1$ values are obtained in multiqubit chips including
complete sets of control wiring.

The improvement of $T_1$ with increasing width is consistent with
previous experiments on superconducting resonators
\cite{barends2010,gao2008}. Loss arises from the electric fields
coupling to two-level systems with dipole moments \cite{phillips},
which reside predominantly in surface oxides and interfaces. This
loss depends on the participation ratio, which depends on the
electric field distribution \cite{wenner2011}. Widening the capacitor
reduces the participation of the surfaces and thus the loss, a
natural explanation for the approximately linear increase in average
$T_1$ with width in Fig.~\ref{fig:t1}. On the other hand, the peaks
in the decay rate are reminiscent of experiments with phase qubits
\cite{martinis2005}, where localized features in the frequency
dependence occur when the qubit couples strongly to two-level
defects, often giving rise to splitting of the qubit frequency and
the chevron-shaped signature of coherent swapping. However, the
exponential decay in the Xmon qubit, with no signatures of swapping
or splitting, suggests a different energy relaxation mechanism.

Here, we show how surface defects near the metal edges of the
capacitor provide a natural explanation for the peaks in the energy
decay. The key point is that loss arises from the qubit interacting
with a sparse bath of incoherent defects, as indicated by our data:
the sharp frequency dependence as well as the changes in fine
structure when cycling to 4.2~K are both clear indicators of defects.
The absence of chevrons and qubit frequency splittings correspond to
incoherent interaction. The lower and upper bounds of $T_1$
increasing with capacitor dimension indicates the defects reside in
the capacitor.

We model a quantum system consisting of a qubit, with a
frequency-independent background loss rate $\Gamma_{1,\mathrm{Q}}$
and pure dephasing rate $\Gamma_{\phi,Q}$, and a single two-level
defect with decoherence rate $\Gamma_{1,\mathrm{D}}$ and dephasing
rate $\Gamma_{\phi,\mathrm{D}}$ (see Supplementary Material
\cite{supmat}). When the defect decoherence rate exceeds the coupling
strength $g$, coherent swapping vanishes and an incoherent,
exponential decay appears. From a two-spin Hamiltonian (see
Supplementary Material), we derive the qubit energy relaxation rate
$\Gamma_1$ (in the limit
$\Gamma_{1,\mathrm{D}}>g>\Gamma_{1,\mathrm{Q}}$) to be
\begin{align}
\label{eq:incoherentdecay}
\Gamma_1 = \frac{2 g^2 \Gamma}{\Gamma^2+ \Delta^2} + \Gamma_{1,\mathrm{Q}},
\end{align}
with detuning $\Delta$, and
$\Gamma=\Gamma_{1,\mathrm{D}}/2+\Gamma_{\phi,\mathrm{D}}+\Gamma_{1,\mathrm{Q}}/2+\Gamma_{\phi,\mathrm{Q}}$.
Hence, each uncorrelated defect adds a single Lorentzian to the
energy decay rate. We can roughly estimate the coupling strength $g$
for a surface defect with dipole moment $p\sim1$~D at a distance $x$
away from the metal edge. With the electric field given by
$E=B/\sqrt{x}$ \cite{jackson}, and $B$ from numerical simulations
\cite{efield}, we arrive at $g/2\pi \sim 0.1$~MHz ($g=pE$) for a
defect located at $x=3$~nm away from the metal. We apply our model to
Fig.~\ref{fig:swapspec}c and find that the peaks in decay rate can be
described by a set of Lorentzians, with
$1/(\Gamma_{1,\mathrm{D}}/2+\Gamma_{\phi,\mathrm{D}})\sim 50-100$~ns,
consistent with defect decay rates measured in similar systems
\cite{martinis2005,shalibo}, and with $g/2\pi \gtrsim0.2$~MHz,
agreeing with incoherent loss.

We can also estimate the number of individually resolvable defects
using two-level system physics developed for junctions. The
substrate-metal interface in our devices was thoroughly cleaned
\cite{megrant2012}, hence we assume that the bulk of strongly-coupled
defects resides in the metal- and substrate-air interfaces, as they
have the highest participation ratios \cite{wenner2011}. The defect
density for AlO$_x$ in tunnel barriers has been established in
measurements with phase qubits \cite{martinis2005}, with the
distribution over dipole moment given by
$\rho_0\sqrt{1-p^2/p_{\mathrm{max}}^2}/p$, with
$\rho_0\approx10^2/\mu$m$^3$/GHz, and the maximum dipole moment
$p_{\mathrm{max}}=6$~D. We take these numbers as representative and
assume a 3~nm thick dielectric layer with defects \cite{thickness}.
The number of defects with coupling strength greater than
$g_{\mathrm{min}}$ is then given by
\begin{align}
N=\iint \rho_0 \frac{\sqrt{1-p^2/p_{\mathrm{max}}^2}}{p} \Theta \left[ p|E(\vec{r})| - g_{\mathrm{min}} \right] dp d\vec{r},
\label{eq:ntls}
\end{align}
with $\Theta$ the unit step function and $E(\vec{r})$ the electric
field at position $\vec{r}$. Simulations using Eq.~\ref{eq:ntls} as
well as Monte Carlo simulations indicate $N\sim 30-50$/GHz, for
$g_{\mathrm{min}}/2\pi\sim0.2$~MHz. We emphasize that the simulations
connect $g_{\mathrm{min}}$ to $N$ with values which are close to what
is observed experimentally. The Monte Carlo simulations (see
Supplementary Material \cite{supmat}) also indicate that the bulk of
strongly coupled defects reside within a $\sim$100~nm distance from
the etched metal edges, where the electric fields are largest. In
addition, we have simulated the qubit decay rate (see Supplementary
Material), reproducing the experimentally observed features: the
simulated decay rate shows both the peaks as well as the variation in
the background.

The good quantitative comparison between model and experiment gives
compelling evidence that a sparse bath of incoherent defects plays a
major role in loss in highly coherent qubits. Our results may also
explain previously reported anomalous behavior in planar transmon
qubits with long coherence, for which the $T_1$ has been reported to
vary significantly between qubits, even on the same chip
\cite{chow,sandberg2012}. This is consistent with a sparse bath of
incoherent defects limiting the coherence, as in
Fig.~\ref{fig:swapspec}.

In conclusion, we demonstrate energy coherence times exceeding
40~$\mu$s in tunable, planar superconducting qubits. We have achieved
this using a geometry with low radiative dissipation and high quality
materials. At these high coherence levels we identify a novel
decoherence mechanism, loss from a sparse bath of incoherent defects,
which is apparent in the swap spectroscopy. We show how enlarging the
capacitor improves coherence. Our qubits combine long coherence, easy
interconnectivity, and fast control, providing a key ingredient for
the implementation of an on-chip surface code quantum computer.

This work was supported by the Rubicon program of the Netherlands
Organisation for Scientific Research (NWO), and the Office of the
Director of National Intelligence (ODNI), Intelligence Advanced
Research Projects Activity (IARPA), through the Army Research Office
grants W911NF-09-1-0375 and W911NF-10-1-0334. All statements of fact,
opinion or conclusions contained herein are those of the authors and
should not be construed as representing the official views or
policies of IARPA, the ODNI, or the U.S. Government. Devices were
made at the UC Santa Barbara Nanofabrication Facility, a part of the
NSF-funded National Nanotechnology Infrastructure Network, and at the
NanoStructures Cleanroom Facility.


\begin{thebibliography}{99}
\bibitem{macc} We measure the effective inductive coupling between
    qubit and inductive driveline \cite{koch} to be $M'=$0.1~pH.
\bibitem{ease} The Xmon capacitance is closely approximated by
    the coplanar waveguide formula: ${C=8L(\epsilon_0+\epsilon_\mathrm{sub})} {K(k)/K(\sqrt{1-k^2})}$, with $\epsilon_0$ the vacuum permittivity, $\epsilon_\mathrm{sub}$ the substrate permittivity, $k=S/(S+2W)$, and $K$ the complete elliptic integral.
\bibitem{efield} For the $S=8$, $W=8$ geometry, we find for the
    electric field away from the metal edge in the plane of the substrate surface:
    $E=B/\sqrt{x}$, with $B=2$~mV$/\sqrt{\mathrm{m}}$.
\bibitem{thickness} Using a thickness of 2~nm instead of 3~nm for the
    dielectric layer does not significantly affect the outcome.
\end{thebibliography}
\end{document}


\title{Supplementary material for ``Coherent Josephson qubit suitable for scalable quantum integrated circuits''}

\author{R. Barends}
\thanks{These authors contributed equally to this work}
\author{J. Kelly}
\thanks{These authors contributed equally to this work}
\author{A. Megrant}
\author{D. Sank}
\author{E. Jeffrey}
\author{Y. Chen}
\author{Y. Yin}
\altaffiliation[Present address: ]{Department of Physics, Zhejiang
University, Hangzhou 310027, China}
\author{B. Chiaro}
\author{J. Mutus}
\author{C. Neill}
\author{P. O'Malley}
\author{P. Roushan}
\author{J. Wenner}
\author{T. C. White}
\author{A. N. Cleland}
\author{John M. Martinis}
\affiliation{Department of Physics, University of California, Santa
Barbara, CA 93106, USA}

\maketitle

\section{Device Fabrication}

The devices were made in a two-step deposition process. The qubit
capacitor, groundplane, readout resonator and control and readout
wiring were made in a first, separate deposition step. We used
molecular beam epitaxy (MBE) Al deposited on a \emph{c}-plane
sapphire substrate. The Al film thickness is approximately 100~nm,
deposited at room temperature. The sapphire substrate was cleaned by
load-lock outgassing at $200\,^{\circ}\mathrm{C}$, followed by
heating to $850\,^{\circ}\mathrm{C}$ in $\sim 10^{-6}$ Torr activated
oxygen, identical to the process outlined in Ref. \cite{megrant2012}.
The first Al layer was patterned by a BCl$_3$/Cl$_2$ reactive ion
etch.

\begin{figure}[b]
    \centering
    \includegraphics[width=0.48\textwidth]{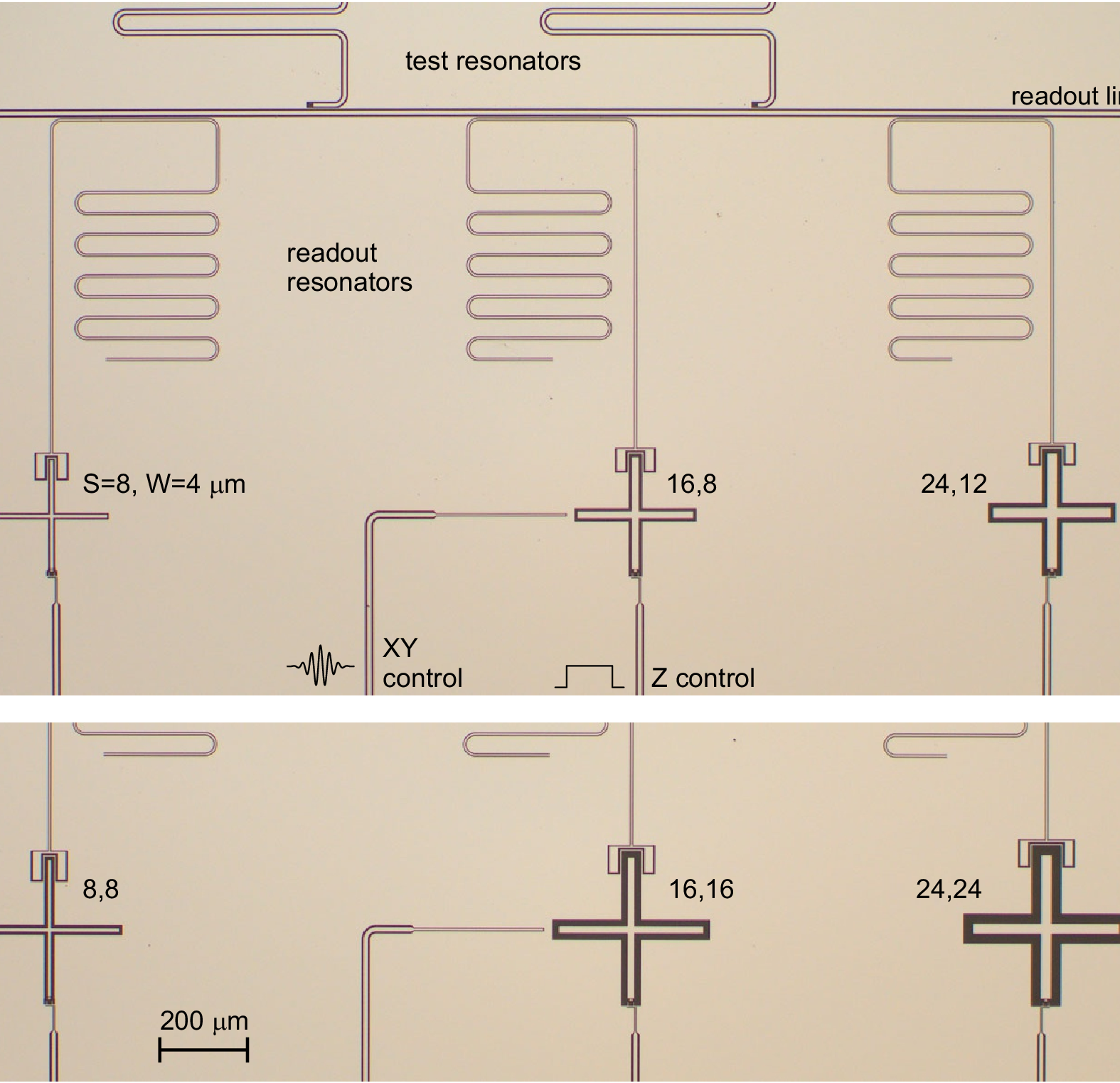}
    \caption{(Color online) Optical micrograph of the six `Xmon' qubits on two chips, formed by the Al superconducting (light) and the exposed sapphire substrate (dark).
    The qubits are capacitively coupled to readout resonators, which couple to a readout line in a frequency multiplexed readout scheme \cite{yu2012}.
    The central linewidth $S$ and gap width $W$ are varied from 8,4 $\mu$m to 24,24 $\mu$m.
    The top three Xmons have a single arm length of $L=130~\mu$m, the bottom three have $L=165~\mu$m.
    Test resonators provide an independent measurement of the quality factor.}
    \label{fig:alldevices}
\end{figure}

In the final step, 0.30 x 0.20 $\mu$m$^2$ Al tunnel barriers (30~nm
bottom and 100~nm top layer thickness) were made using double-angle
shadow evaporation. We used a high vacuum electron-beam evaporator,
with a base pressure of approximately $5\times 10^{-8}$ Torr. We used
the Dolan bridge technique with a poly(methyl methacrylate)/copolymer
resist bilayer (approximate thickness: 0.30 and 0.50~$\mu$m,
respectively), patterned with electron beam lithography. In order to
make galvanic contact between the first Al layer and the junction
layer, we used a 3 min long Ar ion mill (beam: 400~V, 21~mA; beam
width: $\sim$3.2") before shadow-evaporation. Approximately 40~nm was
removed from the top resist layer during the ion mill. The junctions
were oxidized for 80 mins at 5.0 mBar. Lift-off was done in
N-methyl-2-pyrrolidinone at $80\,^{\circ}\mathrm{C}$.

We find that the junctions age very little, the resistance value
changes less than 1\% over a period of ten days. An optical
micrograph of the devices is shown in Fig.~\ref{fig:alldevices}.

\section{Z pulse}

\begin{figure}[b]
    \centering
    \includegraphics[width=0.5\textwidth]{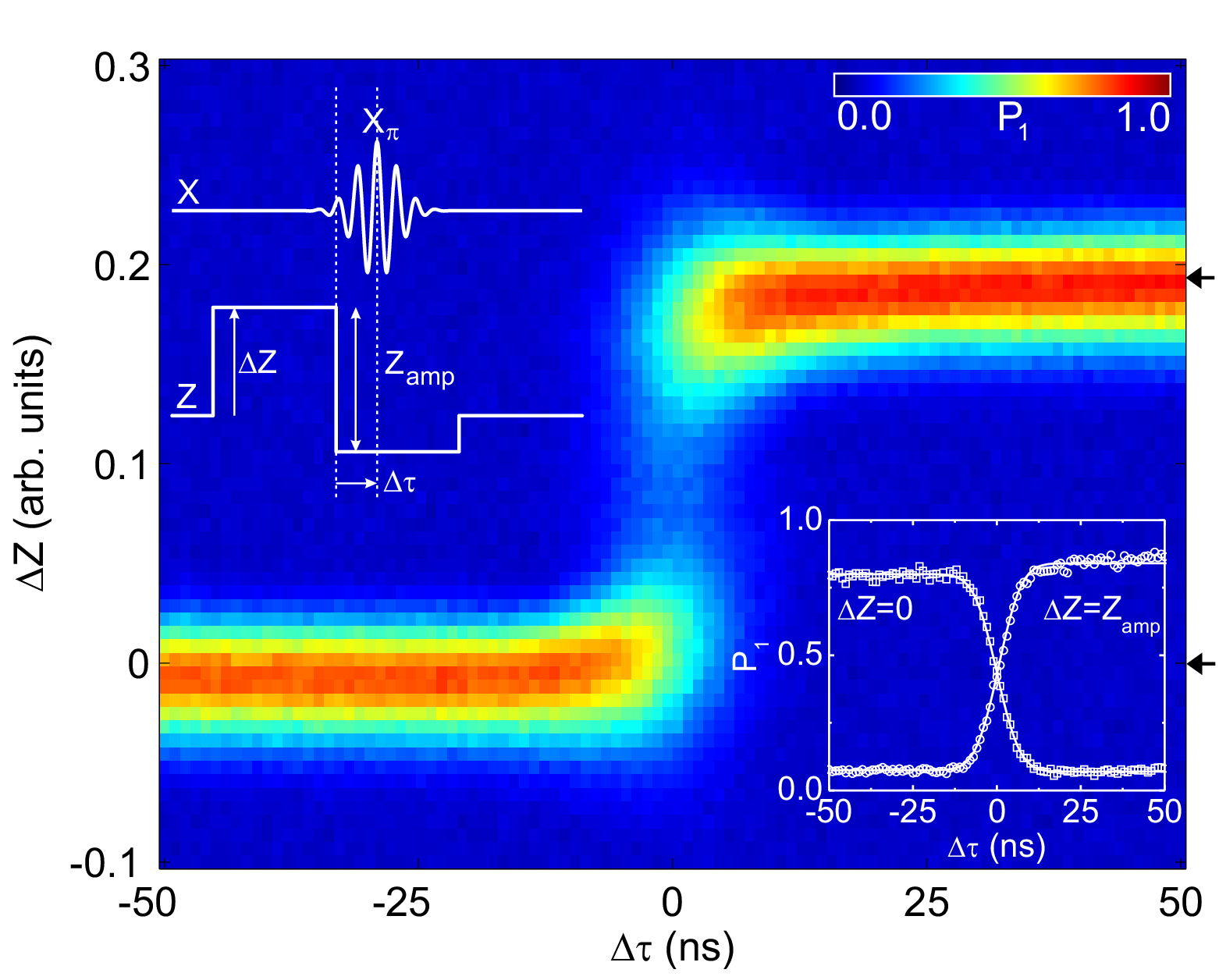}
    \caption{(Color online) Z pulse shape measurement. The qubit is excited with a $\pi$-pulse with 20~ns Gaussian envelope,
    while a rectangular wave is applied on the Z line.
    The pulse sequence is shown in the left inset. We use $Z_{\mathrm{amp}}=0.2$ (arb. units) \cite{zarb}.
    The right inset is a cross section at $\Delta Z=0$ (squares) and $\Delta Z=Z_{\mathrm{amp}}$ (circles),
    as indicated by the arrows (right).
    The solid lines are fits to partial qubit rotations from a $\pi$-pulse with a truncated Gaussian envelope.}
    \label{fig:zpulse}
\end{figure}

In order to quantify the response time of the qubit to a Z pulse, we
simultaneously apply a $\pi$-pulse with a Gaussian envelope on the XY
control and a rectangular wave pattern on the Z control line. The
pulse sequence is shown in the left inset in Fig. \ref{fig:zpulse}.
We slide the rectangular wave in time ($\Delta \tau$) and offset
($\Delta Z$), while retaining the amplitude of the wave
$Z_\mathrm{amp}$ constant. The excited state population is plotted in
Fig. \ref{fig:zpulse} as a function of time and amplitude.

A cross section of the main figure at $\Delta Z=0$ (squares) and
$\Delta Z=Z_{\mathrm{amp}}$ (circles) is shown in the right inset.
The measured response can be accurately described by qubit rotation
from a partial $\pi$-pulse ($\pi$-pulse duration: 20~ns), with a
truncated Gaussian as envelope (solid lines). No other time constants
are included. We find that the rise at $\Delta Z=Z_{\mathrm{amp}}$ is
best described when assuming a $0.5~$ns delay compared to the fall at
$\Delta Z=0$. We conclude that the qubit frequency is tuned to the
desired frequency on a timescale of nanoseconds.

We find the Z control cross-talk between adjacent Xmons to be
1.0-1.5~\%.

\section{Qubit decay rate}

\begin{figure}[b]
    \centering
    \includegraphics[width=0.48\textwidth]{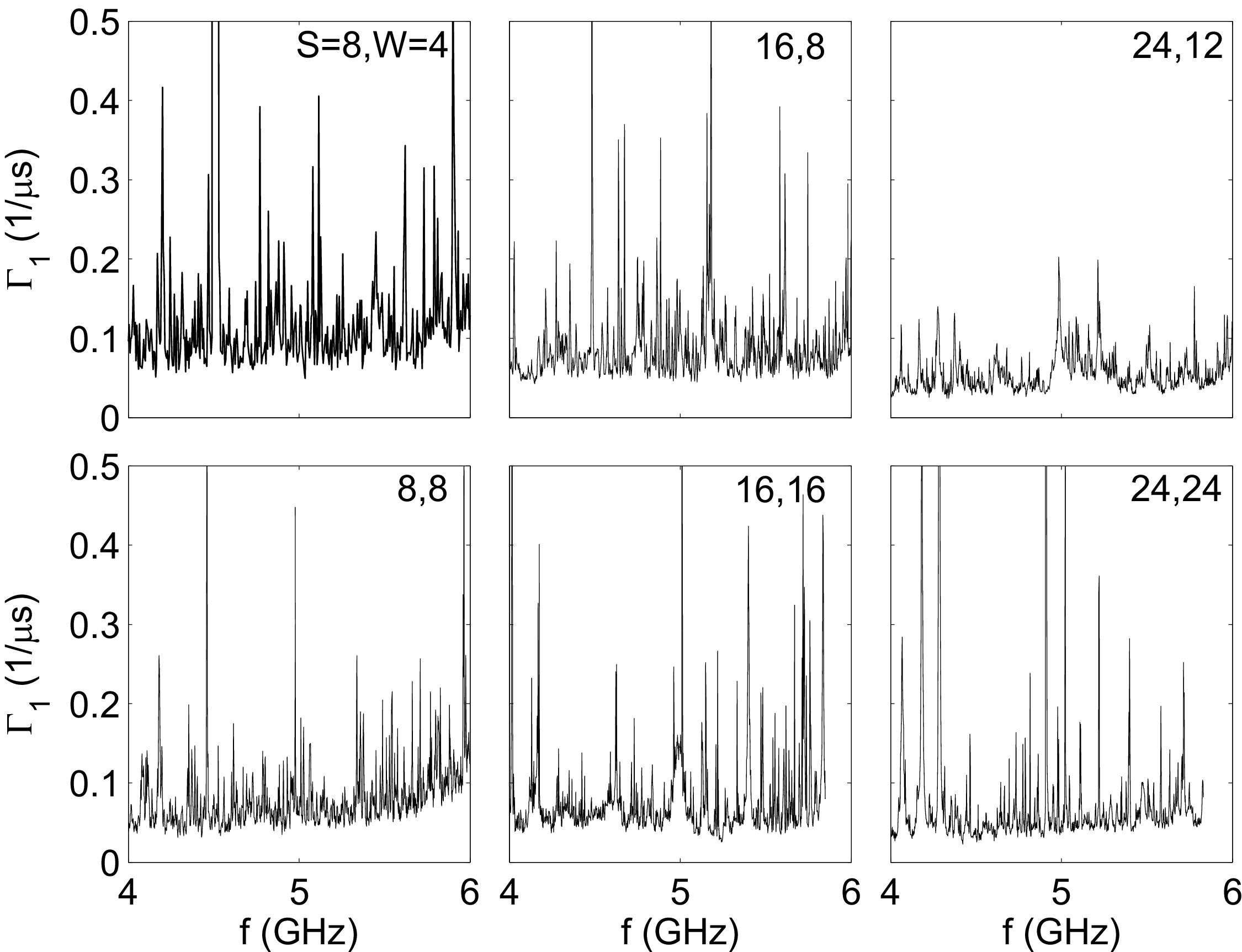}
    \caption{Frequency dependence of $\Gamma_1$ for six qubits with different $S$ and $W$.
    The frequency stepsize is 5~MHz for $S$,$W$ = 8,4~$\mu$m and 2~MHz otherwise.}
    \label{fig:g1}
\end{figure}

The frequency-dependent decay rate for the six qubits, shown in
Fig.~\ref{fig:alldevices}, is displayed in Fig.~\ref{fig:g1}.

\section{Simulation of a qubit-defect system}

\begin{figure}[t]
    \centering
    \includegraphics[width=0.48\textwidth]{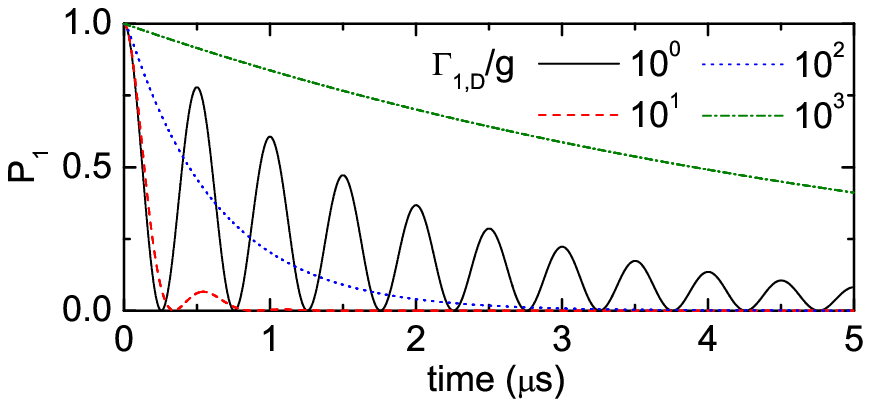}
    \caption{(Color online) Qubit excited state probability versus time, in a simulation of a qubit-defect system.
    The qubit is on resonance with the defect, whose energy relaxation time is $\Gamma_{1,\mathrm{D}}$.
    Here, we used a coupling strength $g/2\pi=1$ MHz.}
    \label{fig:qsim}
\end{figure}

In order to elucidate the crossover from coherent to incoherent decay
of the qubit state, we numerically simulate a qubit-defect system.
The system consists of two coupled two-level systems, with coupling
strength $g$ and defect energy decay rate $\Gamma_{1,\mathrm{D}}$;
the qubit is placed on resonance with the defect. The qubit excited
state probability is shown in Fig.~\ref{fig:qsim}. For
$\Gamma_{1,\mathrm{D}}<g$, the excitation coherently swaps back and
forth between qubit and defect, decaying slowly. When the decay rate
exceeds the coupling strength ($\Gamma_{1,\mathrm{D}}>g$) coherent
swapping vanishes and an exponential decay appears, as the qubit
state decays incoherently. Interestingly, the excitation decays most
quickly for $\Gamma_{1,\mathrm{D}}\approx g$.

\section{Analytical expression for loss in a qubit-defect system}

Here we derive an analytical expression for the energy loss rate
arising from a qubit coupling to a single two-level defect. We
consider a system with two coupled two-level systems. We solve the
master equation in the Lindblad form,
\begin{align}
\dot{\rho}=-\frac{i}{\hbar} [H,\rho] + \sum_{i} \mathcal{D}\left[ C_i \right]\rho,
\end{align}
with $\mathcal{D}\left[ C \right]\rho = C \rho C^+ - \left( C^+ C
\rho + \rho C^+ C \right)/2$ and the Hamiltonian given by
\begin{align}\
\label{eq:hamiltonian}
H = \hbar \omega_\mathrm{Q} a^+ a  + \hbar \omega_\mathrm{D} b^+ b + \hbar g (b^+ \otimes a + b \otimes a^+),
\end{align}
with $a$ and $b$ the lowering operator for qubit and defect,
respectively, and $\omega$ the radial transition frequency. We model
Markovian decoherence through the Lindblad terms: $C_{i=1-4} =
\{a\sqrt{\Gamma_{1,\mathrm{Q}}},~a^+a\sqrt{2\Gamma_{\phi,\mathrm{Q}}}
,~b\sqrt{\Gamma_{1,\mathrm{D}}},~b^+b\sqrt{2\Gamma_{\phi,\mathrm{D}}}\}$,
denoting energy and phase relaxation for qubit and defect,
respectively. Here we used the number operators $a^+a$ and $b^+b$ to
express pure dephasing, and $\Gamma_{1,\mathrm{Q}}$ denotes qubit
relaxation. We take $\Gamma_{1,\mathrm{Q}}\ll\Gamma_{1,\mathrm{D}}$.
As we are interested in the relaxation of a single excitation, we
only consider the states $\{\ket{00},~\ket{01},~\ket{10}\}$. In the
interaction picture and matrix form the above becomes
\begin{align}
H =& \hbar \begin{pmatrix}
0 & 0 & 0 \\
0 & 0 & g \\
0 & g & \Delta \end{pmatrix}, \\ \nonumber
\dot{\rho} = & -\frac{i}{\hbar} [H,\rho]\\ \nonumber
-&
\frac{\Gamma_{1,\mathrm{Q}}}{2}\begin{pmatrix}
-2\rho_{22} & \rho_{12} & 0 \\
\rho_{21} & 2\rho_{22} & \rho_{23} \\
0 & \rho_{32} & 0 \end{pmatrix}
-
\Gamma_{\phi,\mathrm{Q}}\begin{pmatrix}
0 & \rho_{12} & 0 \\
\rho_{21} & 0 & \rho_{23} \\
0 & \rho_{32} & 0 \end{pmatrix}\\
-&
\frac{\Gamma_{1,\mathrm{D}}}{2}\begin{pmatrix}
-2\rho_{33} & 0 & \rho_{13} \\
0 & 0 & \rho_{23} \\
\rho_{31} & \rho_{32} & 2\rho_{33} \end{pmatrix}
-
\Gamma_{\phi,\mathrm{D}}\begin{pmatrix}
0 & 0 & \rho_{13} \\
0 & 0 & \rho_{23} \\
\rho_{31} & \rho_{32} & 0 \end{pmatrix},
\end{align}
with $\rho$ the density matrix, and $\Delta = \omega_T - \omega_Q$.

We are interested in the decay of the qubit excited state probability
$\rho_{22}$, the relevant equations extracted from above are
\begin{subequations}
\begin{align}
\dot{\rho}_{22} &= -i g (\overline{\rho_{23}} - \rho_{23}) - \Gamma_{1,\mathrm{Q}} \rho_{22} \\
\dot{\rho}_{23} &= -i g (\rho_{33} - \rho_{22}) +i\Delta\rho_{23} - \Gamma \rho_{23}  \\
\dot{\rho}_{33} &= -i g  (\rho_{23} - \overline{\rho_{23}})  - \Gamma_{1,\mathrm{D}}\rho_{33}
\end{align}
\end{subequations}
with
$\Gamma=\Gamma_{1,\mathrm{D}}/2+\Gamma_{\phi,\mathrm{D}}+\Gamma_{1,\mathrm{Q}}/2+\Gamma_{\phi,\mathrm{Q}}$.

In the limit $\Gamma_{1,\mathrm{D}} > g$, $\rho_{33}\approx 0$, and
we can approximate the system with two coupled differential
equations. Inserting $\rho_{22}=\exp(-\Gamma_1t)$ and
$\rho_{23}=(\beta_r + i \beta_i)\exp(-\Gamma_1t)$ gives
\begin{subequations}
\begin{align}
\Gamma_1 &= 2g \beta_i + \Gamma_{1,\mathrm{Q}} \\
\Gamma_1 \beta_r &= \Delta \beta_i + \Gamma \beta_r \\
\Gamma_1 \beta_i &= - g - \Delta \beta_r + \Gamma \beta_i,
\end{align}
\end{subequations}
The solution for the qubit energy decay rate $\Gamma_1$ in the
presence of a two-level defect is
(for~$\Gamma_{1,\mathrm{D}}>g>\Gamma_{1,\mathrm{Q}}$)
\begin{align}
\Gamma_1 = \frac{2 g^2 \Gamma}{\Gamma^2+ \Delta^2} + \Gamma_{1,\mathrm{Q}}.
\label{eq:Gamma1}
\end{align}

\section{Monte Carlo simulation of defects in the Xmon qubit}

\begin{figure}[t]
    \centering
    \includegraphics[width=0.48\textwidth]{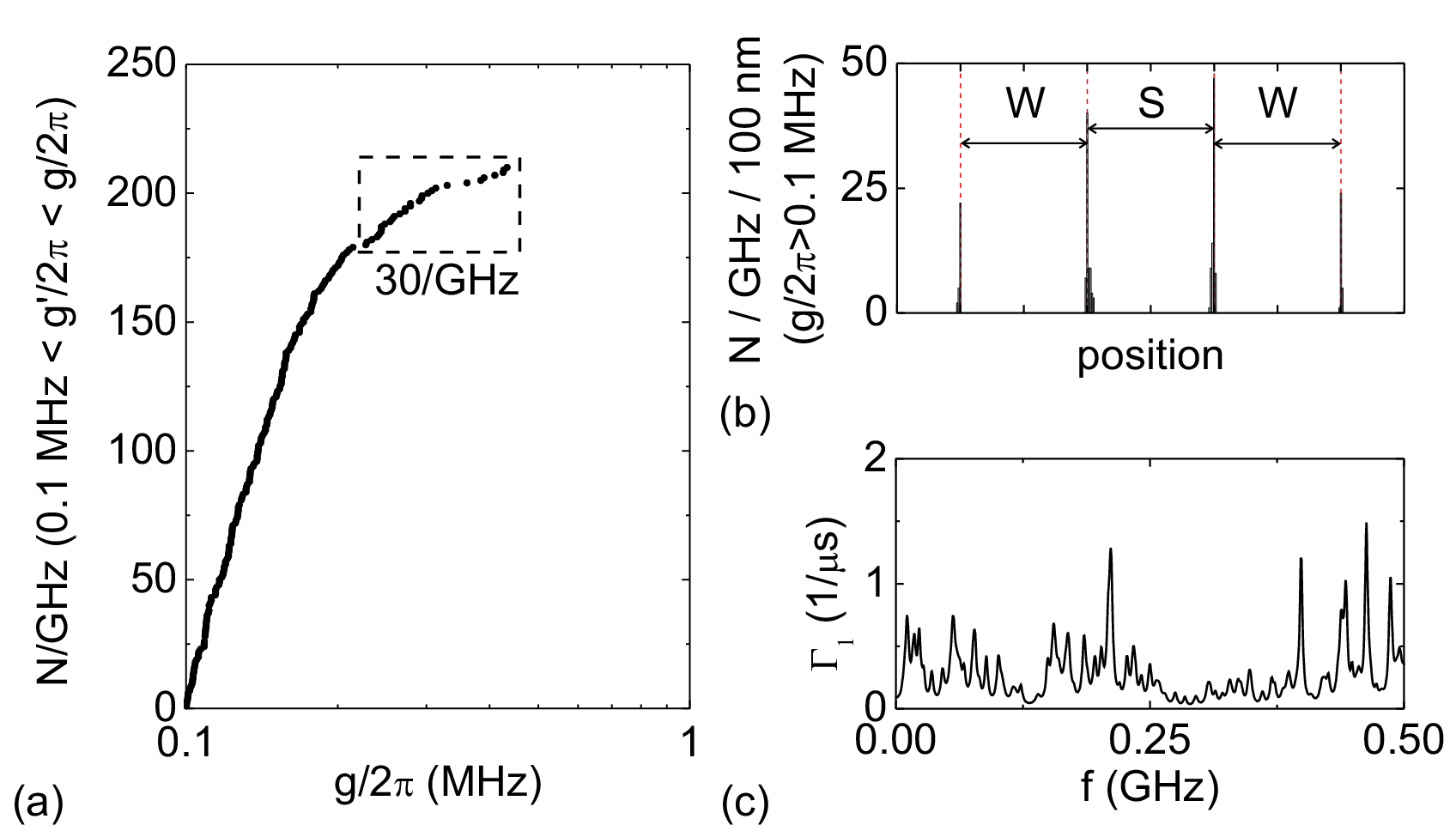}
    \caption{(Color online) Monte Carlo simulation for defects in the Xmon capacitor,
    with $S$=8, $W$=8~$\mu$m and the Xmon arm length $L$=165 $\mu$m.
    (a) Distribution of the defect coupling strength.
    A total of 210 defects/GHz have a coupling strength $g/2\pi>0.1$~MHz.
    A total of 30 defects/GHz have a coupling strength $g/2\pi>0.22$~MHz.
    (b) Distribution of the defects along position; number of defects in each 100~nm wide section is plotted.
    The cross section of the capacitor is dashed.
    (c) The simulated decay rate in a 0.5~GHz bandwidth using the distribution in (a,b), and Eq.~\ref{eq:Gamma1}.}
    \label{fig:montecarlo}
\end{figure}

In order to quantitatively understand the Xmon qubit's energy decay
as well as its variation over frequency, we have performed a Monte
Carlo simulation for defects in the capacitor. The defect density for
AlO$_x$ in tunnel barriers has been established in measurements with
phase qubits \cite{martinis2005}, with the distribution over dipole
moment $p$, volume, and frequency given by
$\rho_0\sqrt{1-p^2/p_{\mathrm{max}}^2}/p$, with
$\rho_0\approx10^2/\mu$m$^3$/GHz, and the maximum dipole moment
$p_{\mathrm{max}}=6$~D. As the capacitor metal oxide and exposed
substrate both consist of Al oxide, we assume that these numbers are
a fair representation of the defect density in the qubit capacitor.

We randomly place defects, with the proper distribution over $p$, in
a 3~nm thick dielectric layer ($\epsilon_r=10$) on the substrate-air
and metal-air interfaces (top metal surface as well as the etched
edges). The substrate-metal interface is assumed to be thoroughly
cleaned \cite{megrant2012} and to contain no significant defect
density. Using a thickness of 2~nm instead of 3~nm for the dielectric
layer does not significantly influence the results. The coupling
strength $g=pE$ is then calculated using a simulation for the
electric fields in our geometry. The results are shown in
Fig.~\ref{fig:montecarlo}.

In Fig.~\ref{fig:montecarlo}a, we have plotted the defect
distribution over coupling strength. We find that, in a 1~GHz band,
30 defects have a coupling strength of $g/2\pi\gtrsim0.2$~MHz
(square). This simulated value is close to the experimentally
observed density of $\sim$30/GHz. These strongly coupled defects are
predominantly located within a $\sim$100~nm distance from the etched
metal edges, including the exposed substrate surface close to the
metal edges and capacitor metal oxide, where the electric fields are
largest, see Fig.~\ref{fig:montecarlo}b.

The simulated qubit decay rate for the same defect distribution is
shown in Fig.~\ref{fig:montecarlo}c in a 0.5~GHz bandwidth. For the
defect decay rate we have assumed a uniform distribution based on the
experimentally determined values: $1/\Gamma_{2,\mathrm{D}}=50-100$~ns
($\Gamma_{2,\mathrm{D}}=\Gamma_{1,\mathrm{D}}/2+\Gamma_{\phi,\mathrm{D}}$).
The simulated decay rate reproduces both the peaks as well as the
variation in the background which are observed in the measurement.